\documentclass[9pt]{iopart}
\usepackage{graphicx}

\usepackage{multirow} 
\usepackage{rotating}
\usepackage{natbib}
\usepackage{doi}
\usepackage{hyperref}
\usepackage{multicol}

\begin{document}

\title[Auger intensity after $^{125}$I decay]{Measurement of the  intensity ratio of Auger and conversion electrons for the electron capture decay of $^{125}$I }

\author{M. Alotiby$^{1,2}$,	I. Greguric$^3$ T. Kib\'{e}di$^4$, B.Q. Lee$^4$ \footnote{Present address: Department of Oncology, Oxford University,  Oxford, UK}, 
	M. Roberts$^3$, A.E. Stuchbery$^4$, Pi Tee$^4$, T. Tornyi$^4$ \footnote{Present address: ATOMKI, Debrecen, Hungary} and M. Vos$^1$ 
}
\address{$^1$ Electronic Materials Engineering, Research School of Physics and Engineering, Australian National University, Canberra, ACT Australia}
\address{$^2$ King Abdulaziz City for Science and Technology, Riyadh, Saudi Arabia}
\address{$^3$ Australian Nuclear Science and Technology Organisation, Lucas Heights, NSW, Australia}
\address{$^4$ Nuclear Physics, Research School of Physics and Engineering, Australian National University, Canberra, ACT, Australia}
\ead{maarten.vos@anu.edu.au, tibor.kibedi@anu.edu.au}
\vspace{10pt}
\begin{indented}
\item[]  February 2018
\end{indented}

\begin{abstract}
Auger electrons emitted after nuclear decay have potential application in targeted cancer therapy. For this purpose it is important to know the Auger electron yield per nuclear decay. In this work we describe a measurement of the ratio of the number of  conversion electrons (emitted as part of the nuclear decay process) to the number of  Auger electrons (emitted as part of the atomic relaxation process after the nuclear decay) for the case of $^{125}$I. Results are compared with Monte-Carlo type simulations of the relaxation cascade using the BrIccEmis code.  Our results indicate that for  $^{125}$I the  calculations based on rates from the Evaluated Atomic Data Library (EADL) underestimate the K Auger yields  by 20\%. \footnote{accepted for publication as a note  in Physics in Medicine and Biology}
\end{abstract}

%
%
%
%
%

\section{Introduction}
Low energy (10-1000 eV) electrons have a very small mean free path (of the order of nm) for
inelastic excitations. The corresponding high linear energy transfer (LET) values are attractive if
one aims to target tumour cells without collateral damage to neighbouring healthy cells.  A convenient source of low energy electrons is Auger
electron emission after  nuclear decay, and their  use in tumour therapy has been discussed
extensively   \citep{Kassis2004,Tavares2010,Cornelissen2010,Rezaee2017}. 

After nuclear decay the atom is often left
with inner-shell vacancies and this  excited state will decay to the ground state by emission
of a number of  Auger electrons and X-rays.  There are usually  few  Auger electrons with  energies of 20-30 keV and a corresponding range of $\approx 10$
$\mu$m. The vast majority  of Auger electrons have much lower energies, below 1 keV down to almost zero energy.  The size of a  normal mammalian cell is $\approx$ 10 $\mu$m, thus the effects of a specific decay are almost always limited to a single cell.   Due to this short range, Auger emitters are expected to be particularly effective when they are located in the nucleus of a tumour cell as then the probability of double-strand breaking of the DNA is very high, preventing the cell from multiplying \citep{Falzone2017}.


To exploit  their use in nuclear medicine it is thus imperative to have precise
knowledge of the full  energy spectrum of the Auger electrons emitted per nuclear decay.  Atomic relaxation (the return of an atom with an inner core hole to its ground state)   is a
complex process with many possible pathways, especially for higher atomic numbers.  The problem is
most conveniently tackled using  Monte Carlo simulations based on decay rates as calculated for 
isolated atoms \citep{Pomplun1987,Stepanek2000,Nikjoo2008,Lee2016}.  Experimental verification of such results, i.e.
the predicted number of Auger electrons produced per nuclear decay is then highly desirable. 

The two nuclear decay processes producing inner-shell vacancies are electron capture and internal conversion.  The probability of internal conversion involving inner shell electrons is usually  known
within a  percent \citep{Kibedi2008}. 
By comparing the conversion electron (CE) and Auger intensity one can benchmark the
Monte Carlo simulations. Such is the aim of this paper.

$^{125}$I was chosen for the following reasons.  It is one of the most extensively studied radioisotope owing to its possible application to cancer therapy  \citep{Balagurumoorthy2012}. Very recently, in combination with gold nanorparticles, $^{125}$I  was used for targeted imaging and radionuclide therapy \citep{Clanton2018}. In the present study we used $^{125}$I  to measure  the ratio of Auger to conversion electrons.  $^{125}$I decays with a half-life of 59.5 days via electron capture to an excited state of the
$^{125}$Te nucleus. This excited state decays in 93\% of the cases to its ground state by the emission of a  CE.
The half-life of this excited state  is 1.48 ns, which is much longer than the time scale for atomic relaxation (femtoseconds).
There are thus two separate relaxation cascades contributing   to the Auger yield: one  after electron capture and the other  after emission of a CE.
The combined large Auger yield makes  $^{125}$I an attractive candidate for targeted tumour therapy.

\section{Experimental Details}
$^{125}$I  can be prepared as a  sub-monolayer source on a Au(111) surface which is
stable in air \citep{Huang1997} and the Te atoms, produced in the decay
process, are bound to this surface as well \citep{Pronschinske2016}. 
Samples with a third of a monolayer of $^{125}$I on a Au(111) surface were prepared as described by \cite{Pronschinske2015}.  Au(111) surfaces were obtained by flame annealing
Au samples (Arrandee Metal GMBH, Germany) just before the $^{125}$I deposition. A droplet containing  NaI in a NaOH solution (pH $\approx 10$, Perkin Elmer) was put on this surface, and left to react. An approximately  4~mm diameter source
 with a strength of 4~MBq  was obtained.

The measurements were performed with two spectrometers.  For  energies below 4 keV (LMM Auger and K CE ), 
the DESA100 SuperCMA  (Staib Instruments) was used.
The spectrometer  was slightly modified by incorporating
high-$Z$ metal shielding to prevent X-rays emitted by the sample interacting with the channeltron detector.
The second spectrometer was  locally-built and measures electrons with higher energies up to 40 keV (the KLL Auger and L CE)
 \citep{Vos00,Went2005}. This spectrometer has a smaller opening angle, but it
is equipped with a two-dimensional detector, measuring a range  of energies simultaneously  (17\% of the pass
energy of 1 keV).  For both spectrometers the Auger signal is on top of a background due to the dark count rate of the detector, which did not depend on the  electron  energy being measured.

As the Auger energies  are different from the CE energies, it is essential to understand
how the  spectrometers efficiency varies with the electron energy $E$.  The  efficiency  of the DESA100 was determined experimentally before \citep{Gergeley1999} and
for energies higher than a few times its pass energy, those results  indicate an efficiency that scales as $1/E^{1.2}$.   Our  SIMION electron optics
simulations \citep{Dahl2000} suggest a somewhat weaker dependence (proportional to  $\approx 1/E^{0.8}$). In the analysis presented here a simple $1/E$ dependence was used. As the main LMM Auger line energies differ by less than 20\% from the K-CE electron energy the CE to Auger intensity ratio is only affected on a 5\% level if a  $ 1/E^{0.8}$ or a $1/E^{1.2}$ dependence is assumed instead of a simple $1/E$ dependence.
The high-energy spectrometer  uses a lens stack  behind a   0.5 mm wide  slit.  It decelerates the electrons and focuses them at 
the entrance of an hemispherical analyser. SIMION simulations showed that
all electrons transmitted through the slit  will enter the analyser.  The spectrometer transmission 
is thus determined solely by the width of the entrance slit and  is independent of $E$. The fact that the  L-shell CEs have $\approx$ 30\% higher energy than
the KLL Auger electrons should thus not affect the comparison of their intensities.

\section{Results}

\begin{figure}
	\centering
	\includegraphics[width=1.0\linewidth]{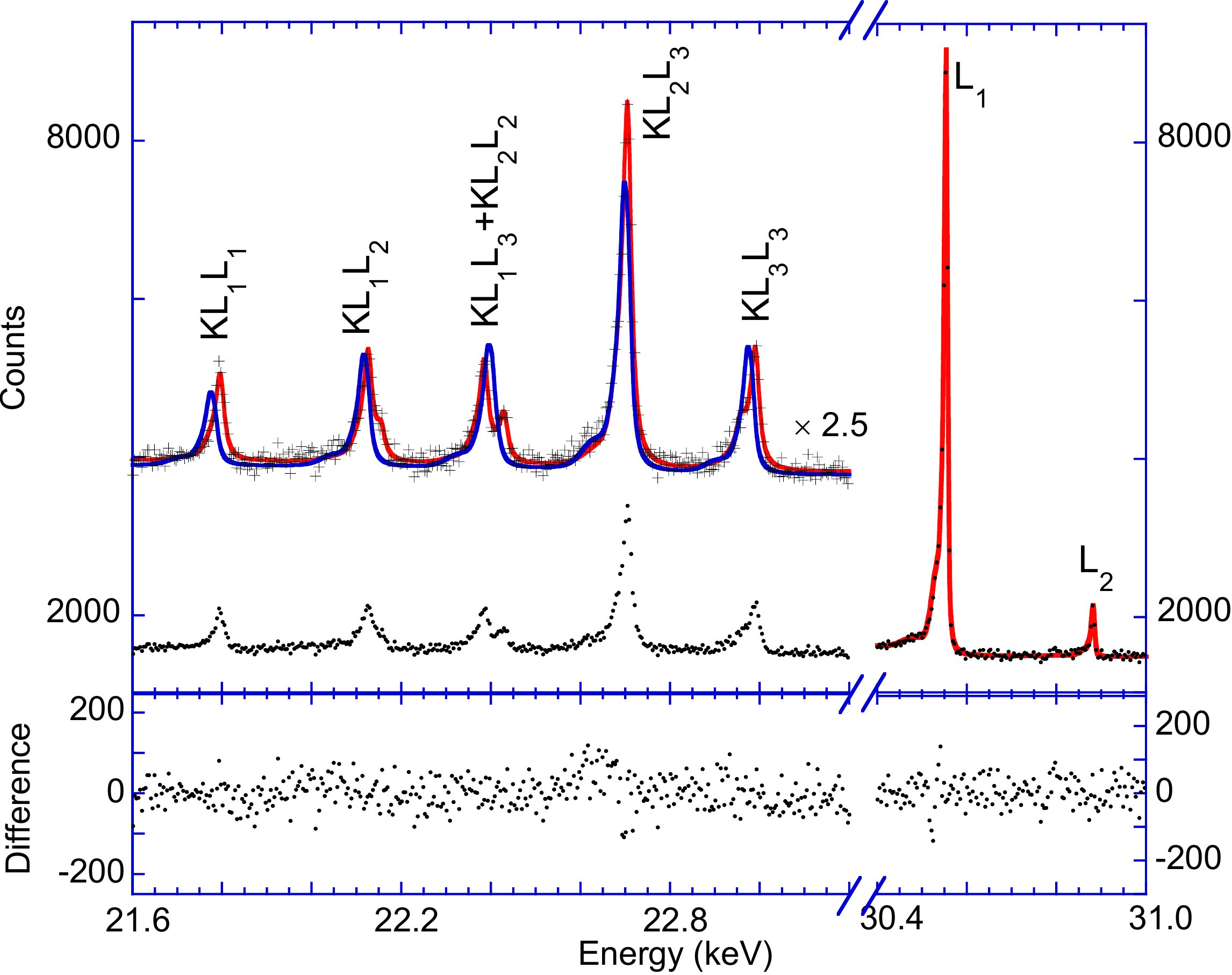}
	\caption{The KLL Auger and L$_1$, L$_2$-CE spectrum as measured in a single run.  The red line shows a fit of the KLL spectrum with 8 peaks following the approach of \cite{Larkins1977} (residuals in lower panel).    The blue line shows a description of the KLL spectrum based on the BrIccEmis calculation\citep{Lee2016} which was scaled such that the calculated L$_1$ CE line has the same area as  the measured one. The calculated Auger spectrum, normalised in this way, has an area that is smaller than the observed one.   }
	\label{fig:klll2fig}
\end{figure}
\subsection{High-Energy  Auger}
 The  KLL Auger spectrum together with the  L$_1$ and L$_2$-CE line is shown in Fig. \ref{fig:klll2fig}.  The Auger part of the spectrum is similar to those
obtained with a magnetic spectrometer by \cite{Graham1962}.  The KLL Auger spectrum consists of several peaks. There are (at least) two ways to describe this
spectrum:  \\
(i) One can characterize each final state in terms of the atomic orbitals they originate from and to the total
angular momentum and total spin quantum number of the final state.  This leads to 9 possible final
states in the intermediate coupling scheme. This approach was followed by \cite{Larkins1977}
and works well for two core holes, but becomes  cumbersome when more vacancies are present, later
in the cascade.   \\
(ii) One can neglect the fine splitting   and characterize the final state
in terms of L$_{1,2,3}$ only.  Then there are 6  possible final states 
 but for Te the 	KL$_1$L$_3$ and KL$_2$L$_2$ energies are almost
identical and experimentally not  resolved.  This approach is adopted in  BrIccEmis \citep{Lee2016} and remains manageable
when one calculates  several steps down in the relaxation cascade, when more vacancies are present.


A comparison was made with the peak positions as calculated by \cite{Larkins1977} and the
intensity as calculated  by  \cite{Chen1980}, as shown in Fig. \ref{fig:klll2fig} as well. The lines were slightly asymmetric and each line was fitted with 4 Gaussians  and a very small Shirley-type background \citep{Shirley1972}.  The parameters used for this fit i.e. the energy offset (relative to the main peak),  width and relative amplitude were the same for all lines (Auger and CE).   The sum of the four Gaussians was convoluted with a Lorentzian, representing the lifetime broadening. For the Auger peak the lifetime broadening was taken to be the sum of the lifetime broadening of the K level and two L levels, for the CE peak just the L lifetime broadening was included.  K and L lifetimes were taken from
\cite{Krause1979a}. There were clearly 8 different components  in the experimental KLL Auger spectrum. In some cases the calculated energies
were separated by less than the peak width (determined mainly by lifetime broadening) and these
components were taken together. The energy separation of the different components was within 10 eV of those calculated by \cite{Larkins1977} and the relative intensity  of the different components was close (within 3\% of the intensity of the KL$_2$L$_3$ component) to those calculated by \cite{Chen1980}.  The intensity  ratio of the L$_1$ CE line to the KL$_2$L$_3$ Auger line obtained from this fit was $1:0.61\pm 0.01$. 

The BrIccEmis program \citep{Lee2016} was used to describe the data.  It uses nuclear decay data from ENSDF (https://www.nndc.bnl.gov/ensdf/), electron capture rates \citep{Schoenfeld1998}, theoretical  conversion coefficients \citep{Kibedi2008}  and atomic transition rates from EADL \citep{Perkins1991}.  The L$_1$ CE to KL$_2$L$_3$ Auger intensity ratio, as calculated by BrIccEmis, is 1: 0.53 which is clearly lower than the experimentally observed one. 

\subsection {Low-Energy Auger} 
\begin{figure}
	\centering
	\includegraphics[width=0.9\linewidth]{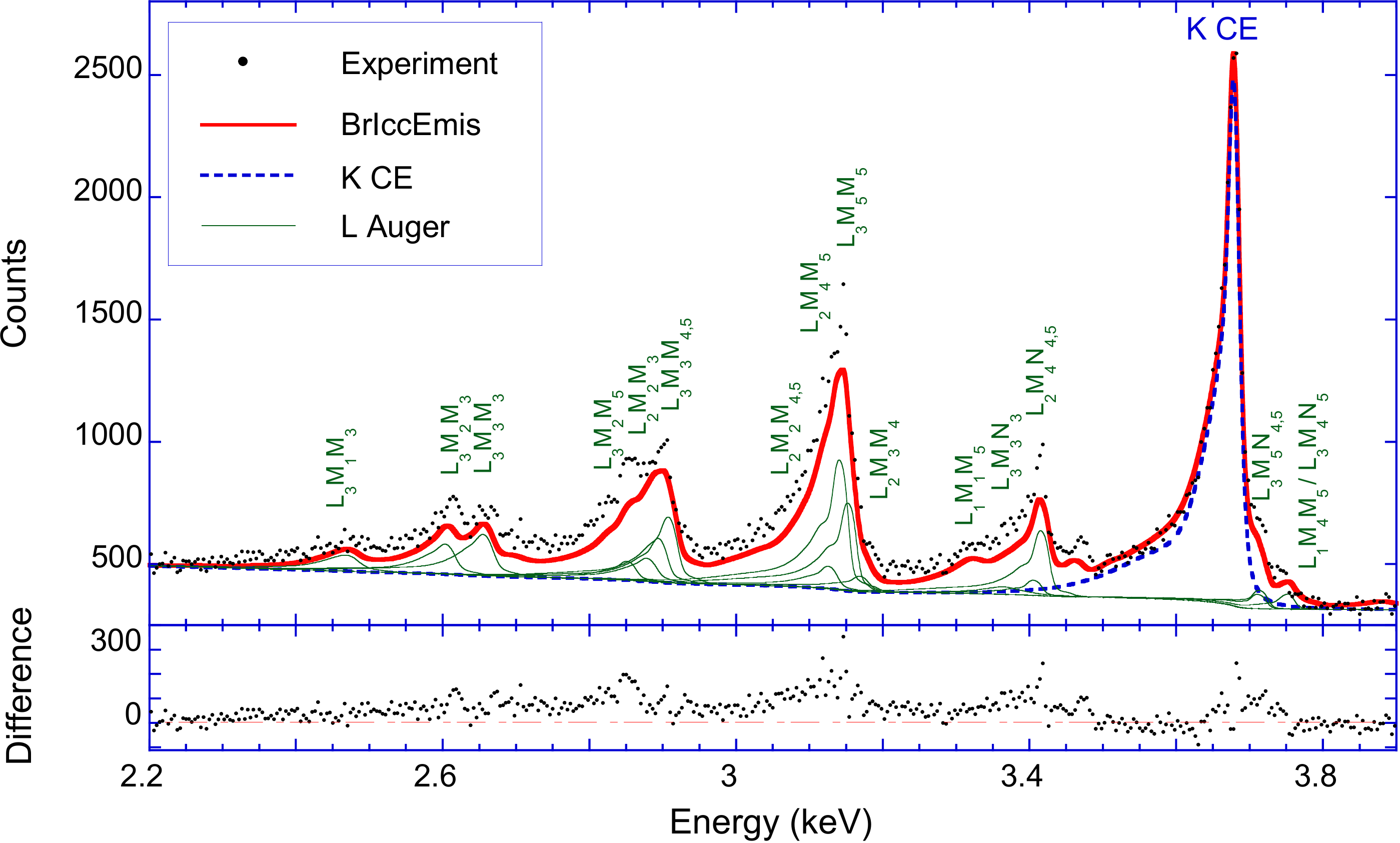}
	\caption{The measured spectrum (dots) of the K-CE and LMM Auger electrons. 
		 The solid red line is the calculated spectrum based
		on BrIccEmis scaled to the K-CE line. The contribution of the conversion
		electrons (blue, dashed line) and the strongest individual Auger electron contributions (thin green lines) are indicated as well.  The lower panel shows the residual of the fit and the non-zero difference indicate that the theory underestimates the Auger intensity, relative to the CE intensity. }
	\label{fig:kconvlmm}
\end{figure}

The K CEs have an energy of 3.679 keV.  This energy is within the range of the LMM
Auger transitions and one can again measure  the ratio of
CE and Auger intensities  experimentally. A spectrum is shown in Fig. \ref{fig:kconvlmm}, and
the peaks are somewhat sharper than those obtained by \cite{Casey1969}  using a magnetic
spectrometer.  The
strongest line is the K-CE line.  There is some overlap between this line and neighbouring
Auger lines. This close proximity makes corrections due to the energy dependence of the analyser
efficiency small.  It, however, makes it more difficult
to assess the exact line shape.  Clearly there is again a tail at the low energy side, but the K
CE line is  broader than the L$_1$-CE one, shown in Fig \ref{fig:klll2fig}.  This is at least in part due to the
larger lifetime broadening of the K hole ($\approx 9.6$ eV versus 3.32  eV for the L$_1$ hole  \citep{Krause1979a}).

An attempt was made to fit the K conversion spectra with the same line shape as the L$_1$ conversion line
(taking into account their different lifetime broadening), but this approach was unsuccessful, likely due to stronger interactions of the lower-energy electrons with their
environment.
Adjusting the line shape by increasing the intensity extending to lower energies (the tail)  to get a better description, and using the same tail for the K CE line and nearby Auger lines 
(and a theoretical estimate of their  lifetimes), we obtain the description of the spectrum based on BrIccEmis shown in
Fig. \ref{fig:kconvlmm}.  Theory was scaled so that the area of the K-CE line was the same as the experimental K-CE peak area.
 The Shirley-type background  at lower energies is now more pronounced,
indicating that the lower energy electrons interact more heavily with the substrate.  However, this procedure  showed that the Auger intensity (relative to the K-CE intensity) was underestimated by $\approx 20$\% in the BrIccEmis calculation. 
This ratio is affected by uncertainties in the spectrometer response and by the procedures used to specify the line shape.  Adjusting the line shape by extending the tail up to 300 eV below the main line 
(and reducing the contribution of the Shirley background at the same time) improves the agreement somewhat, 
but  the calculated Auger intensity remains at least 10\% lower than the experiment.

\section{Conclusion and Discussion}

The combined  K CE - LMM Auger measurement  indicates that the experimental relative Auger intensity is about 15-20\%\   higher than the calculated one.  The same order of magnitude of  difference was found for the  KLL Auger intensity compared to the L$_1$-CE intensity, in spite
of the fact that the energies involved were rather different and that two different spectrometers
were used. 

A core hole can either decay by  X-ray emission (fluorescence) or by Auger decay. The K fluorescence yield, $\omega_K$ is defined as the fraction of K core holes that relaxes by X-ray emission. 
For the  Te K shell, the adopted value based on experimental data
 $\omega_K$ is $\approx 87.5$\% \citep{Hubbell1994,Krause1979}. The EADL database used by BrIccEmis \citep{Perkins1991} uses a very similar value (87.9\%). Some
experimental values for Te are considerably smaller (82.3$\pm 7.3$\%,
\citep{Singh1990}).  
The K-shell Auger yield is equal to (1- $\omega_K$).
 The corresponding Auger yield for the K shell based on theory would be 12.5\%,  whereas  the results of \cite{Singh1990}  correspond to an Auger yield of  17.7\%,  i.e. the
measured fluorescence value from Singh would predict a 50\% larger Auger yield, a difference much 
larger than  what is required to describe our data.  There is thus no experimental evidence that
excludes the possibility that $\omega_K \approx 85$\%, which would describe our data  well. For high $Z$ elements, where $\omega_k$ approaches 1, the determination of the K Auger yield is thus  an accurate way of determining  the value of  $\omega_K$.

\begin{table}
	\begin{center}
	\begin{tabular}{c|cccc|} 
			\cline{2-5} 
 	 &  \multicolumn{4}{c|}{Auger electrons per decay}\\	
 	 \cline{1-5}
		\multicolumn{1}{|r|}{Transition} 			& BrIccEmis & BrIccEmis & Stepanek & Pomplum \\
				\multicolumn{1}{|r|}{  } 	&& mod. $\omega_K$&&\\
		\hline
		\multicolumn{1}{|r|}{KLL}       & 0.130     & 0.155                   & 0.126            & 0.134   \\
		\multicolumn{1}{|r|}{KXY}         & 0.194     & 0.232                   & 0.189            & 0.196   \\
		\multicolumn{1}{|r|}{LMM}       & 1.22      & 1.24                    & 1.22             & 1.25    \\
		\multicolumn{1}{|r|}{LXY}         & 1.86      & 1.89                    & 1.83            & 1.88  \\
		\hline
	\end{tabular}
\caption{The calculated K and L  Auger transition rates based on BrIccEmis with the EADL $\omega_K$ value, with  a modified $\omega_K$ value  to reproduce the experimental KLL intensity and those obtained in the literature by  \cite{Stepanek2000} 
	and \cite{Pomplun1987}. Only the K Auger line intensity depends critically on the $\omega_K$ value used. }
\label{table_yield}
\end{center}
\end{table}

Besides the aforementioned fluorescence-yield measurements based on  results from stable Te isotopes, there are earlier measurements based on coincidences between  $\gamma$ and X-rays for the case of   the decay of $^{125}$I \citep{Karttunen1969}, which gave a value of the fluorescence rate of $85.9\pm 2.2$\%.  It worth noting that the measurement described  here, based exclusively on the measurement of electron intensities,  agrees with the measurement of \cite{Karttunen1969}, which relies solely on  X- and $\gamma$-ray intensities.  

As the L fluorescence yield is low for Te (9\%,  \cite{Hubbell1994}), the LMM Auger intensity is not very
 sensitive to the fluorescence yield. The discrepancy seen for the LMM Auger-K CE intensity ratio  can thus not
be attributed to uncertainty in this quantity. The LMM Auger is generally the second step in the
relaxation cascade, and hence its calculated intensity depends on the processes involved in the
first step, e.g., on how the vacancies are distributed over the L$_1$, L$_2$ and L$_3$ shells after the first relaxation step. Moreover the interpretation is hampered both by limited knowledge of the line shapes involved and the exact dependence of the transmission efficiency of the DESA100. Although the analysis here, based on the assumption of identical  shape of the K CE and LMM Auger lines, indicates that the LMM Auger intensity (relative to the K CE line) is 10-20\% larger than BrIccEmis calculates, it is conceivable that a better understanding of line shapes involved   would resolve this issue.  

In Table \ref{table_yield} we show the calculated Auger intensities for the K and L initial states per nuclear decay using the  BrIccEmis and some calculations from the literature. There is generally a fairly good agreement between the BrIccEmis results and the literature data. In the case of BrIccEmis we used the EADL  $\omega_K$ value of 87.9\% as well as a modified $\omega_K$ value of 85.4\% which is required to fit the experimental KLL Auger intensity.  From this table it is clear that only the K Auger lines are strongly affected by the precise value of the fluorescence rate, whereas the L Auger line intensity is only affected in a very minor way. For medical applications this means that changing the $\omega_K$  value from 87.9\% to 85.4\%   increases the effect of Auger decay   microns away from the emitter by $\approx 20$\%, but at smaller distances (smaller than  the range of LMM Auger electrons $\approx 100$ nm) the effects remain largely  unchanged. 

More generally, we have shown that a comparison of the CE and Auger electron intensity after nuclear
decay provides a crucial test of the theory and thus a clear  way to improve  databases, such as the EADL  by \cite{Perkins1991}, that are widely used in simulating the effects of ionizing radiation in medical physics.

\section{Acknowledgments}
This research was made possible by  Discovery Grant DP140103317 of the Australian Research Council

\newcommand{\newblock}{}

\end{document}